\documentclass[showpacs,aps,twocolumn]{revtex4}
\usepackage{epsfig}
\usepackage{graphicx}
\usepackage{amsmath,amssymb,amsfonts}
\usepackage{array}
\usepackage{url}
\usepackage{hyperref}
\usepackage{multirow}
\usepackage{float}
\usepackage{lineno}
\usepackage{xspace}
\usepackage[usenames,dvipsnames]{color}

\newcommand{\bef}{\begin{figure}}
\newcommand{\eef}{\end{figure}}
\newcommand{\bc}{\begin{center}}
\newcommand{\ec}{\end{center}}

\newcommand{\be}{\begin{equation}}
\newcommand{\ee}{\end{equation}}
\newcommand{\bea}{\begin{eqnarray}}
\newcommand{\eea}{\end{eqnarray}}

\def\ba{\begin{eqnarray}}
\def\ea{\end{eqnarray}}
\begin{document}

\title{Bose-Einstein condensation of pions in proton-proton collisions at the Large Hadron Collider using non-extensive Tsallis statistics }
%\title{Is Bose-Einstein Condensation of pions in Proton-Proton collisions at the Large Hadron Collider possible?}
\author{Suman Deb}
\author{Dushmanta Sahu}
\author{Raghunath Sahoo\footnote{Presently CERN Scientific Associate at CERN, Geneva, Switzerland}}
\email{Raghunath.Sahoo@cern.ch}
\author{Anil Kumar Pradhan}
\affiliation{Department of Physics, Indian Institute of Technology Indore, Simrol, Indore 453552, India}

\begin{abstract}
%Recent LHC results, such as long-range angular correlation, flow-like patterns, strangeness enhancement, etc. in $pp$ collisions, have drawn considerable attention in the study of small systems. This growing interest in studying $pp$ collisions at LHC has led us to further explore various directions. 

The possibility of formation of Bose-Einstein Condensation (BEC) is studied in $pp$ collisions at $\sqrt s$ = 7 TeV at the Large Hadron Collider. A thermodynamically consistent non-extensive formulation of the identified hadron transverse momentum distributions is used to estimate the critical temperature required to form BEC of charged pions, which are the most abundant species in a multi-particle production process in hadronic and nuclear collisions. The obtained results have been contrasted with the systems produced in Pb-Pb collisions to have a better understanding. We observe an explicit dependency of BEC critical temperature and number of particles in the pion condensates on the non-extensive parameter $q$, which is a measure of degree of non-equilibrium -- as $q$ decreases, the critical temperature increases and approaches to the critical temperature obtained from Bose-Einstein statistics without non-extensivity. Studies are performed on the final state multiplicity dependence of number of particles in the pion condensates in a wide range of multiplicity covering hadronic and heavy-ion collisions, using the inputs from experimental transverse momentum spectra.
\end{abstract}
\date{\today}
\maketitle

\section{Introduction}
\label{intro}
In 1924 with the combined efforts of Satyendra Nath Bose and Albert Einstein, a new phenomenon was discovered. In his work, Bose had treated photons as particles of an ideal gas and then he had derived the Planck's law of blackbody radiation with this assumption~\cite{bose}. Later, Einstein predicted that at very low temperatures those particles would condense into the minimum energy level of the system under consideration~\cite{Einstein:1924,P_rez_2010}. This phenomenon is called Bose-Einstein condensation (BEC) and is applicable to particles called bosons, which have integral spins and follow the Bose-Einstein (BE) statistics. BEC is called the fifth state of matter which is usually formed when a system of bosons at low densities are cooled to temperatures very close to absolute zero. Under such conditions, a large fraction of the particles occupy the ground state, where, the wave-functions of the particles interfere with each other and the effect is observed macroscopically. This is possible because of the unique property of bosons whose total wave-functions are always symmetric. So a large number of bosons can occupy the same state unlike the fermions which obey Pauli's exclusion principle. The caveat here is that, BEC usually occurs at very low densities and very low temperatures. However in high energy collisions, the temperature becomes extremely high. In such conditions, whether we can observe any such condensation is a question to be addressed through extensive theoretical studies confronted to appropriate experimentation. The temperature reaches upto hundreds of MeV in the high energy collisions, which is about $10^{10}$ K and is astronomically higher than the temperature required for BEC for cold atoms. But the properties of a BEC for pions would be very different from the low temperature BEC. Firstly, the pions would have much smaller system volume, much higher density and different interactions are also involved in the formation of high temperature BEC \cite{Begun:2015yco}. In an expanding system formed in ultra-relativistic collisions, the temperature will be extremely high as compared to when BEC occurs in systems which are static in nature like helium~\cite{Guha:2018}. It is evident that  most of the particles will have zero momentum at the low temperature BEC. However, in a violently expanding system, the particles will not have zero momentum, but certainly they can have relative momentum close to zero in the phase space. As a basic requirement of BEC formation, the constituents of BEC should be close by both in momentum and configuration space, which is possible in a strongly interacting, highly dense QCD medium. This motivates us to study BEC formation even in expanding medium created in ultra-relativistic collisions such as $pp$ at LHC.
The possibility of observing such a high-temperature BEC of charged pions in hadronic and heavy-ion collisions at the energy and luminosity frontiers in the Large Hadron Collider would be of great interest to the community. 
In this direction, attempts of studying the influence of a possible coherent component in the boson source using chaoticity parameter for the formation of BE correlation were made both theoretically in Refs.\cite{Plumer:1992au,Csorgo:1998tn,Akkelin:2001nd} and experimentally in Refs. \cite{Csanad:2005nr,Gangadharan:2015ina,Adam:2015pbc,Novak:2018hqd}.

The aim of relativistic heavy-ion collisions is to understand the phases of Quantum Chromodynamics (QCD)~\cite{Shuryak:1980tp,Asakawa:1989bq}. In particular, these collisions give us an opportunity to create and characterize a possible deconfined state of quarks and gluons, called Quark-Gluon Plasma (QGP)~\cite{Westfall:1976fu}, which  is believed to have existed after a few micro-seconds of the  Big Bang.  Information about these phases formed in such collisions is extracted from the distribution of the produced particles in the phase space, called (identified)particle spectra. Interpretations of these results are performed using different theoretical models to draw conclusions about the properties of the matter formed at the extremes of temperature and energy density. Statistical hadron gas models (thermal models) and hydrodynamic models~\cite{Hirano:2008hy,Baier:2006um,Teaney:2009qa} are used for this purpose, along with many other variants to confront to the experimental observations.  Hydrodynamics calculations are brought in to correctly explain the anisotropy of the particles originated in the QGP~\cite{Hirano:2008hy}. Thermal models mainly assume thermodynamic equilibrium to explain the hadronic yields and extract the chemical freeze-out parameters like temperature and baryochemical potential. However everything is not so smooth, mainly in explaining the proton to pion ratio and mean multiplicity of proton/anti-proton. Also hydrodynamic models which use local thermodynamic equilibrium, are not quite good at explaining the very low momentum part of the pion transverse momentum ($p_{\rm T}$) spectra \cite{Abelev:2013vea,Molnar:2014zha}. 

To explain these, chemical non-equilibrium in the formation of the hadronic matter is assumed in some studies \cite{Begun:2015ifa}. This brings up the non-zero value of chemical potential which is close to the value of chemical potential needed for BEC of charged pions. These have been the cases studied in the heavy-ion collisions. However, any contribution of non-equilibrium can be well preserved in small systems like that are formed in $pp$ collisions, because of the high gluon density and net-baryon number being negligibly small \cite{Busza:2018rrf}. In such $pp$ collisions at the LHC energies, the baryon chemical potential is close to zero . Moreover, the systems formed in $pp$ collisions are taken as reference to interpret the results of heavy-ion collisions. So it is important to understand the formation of BEC-like features in small systems formed in $pp$ collisions. To this end, we investigate the possibility of BEC in pion gas formed in $pp$ collisions and compare the results with heavy-ion collisions, to find a bridge between the two systems. Such a study of investigating various
phenomena in LHC $pp$ collisions has become a necessity in order to understand the QGP-like features seen in these hadronic collisions~\cite{nature,Li:2011mp,Khachatryan:2010gv}.

It is observed that at RHIC \cite{Abelev:2006cs,Adare:2011vy} and LHC \cite{Aamodt:2011zj,Abelev:2012cn,Abelev:2012jp,Chatrchyan:2012qb} energies, the $p_{\rm T}$-spectra in $pp$ collisions deviate from the standard thermalized Boltzmann-Gibbs (BG) distribution. In such cases, Tsallis non-extensive distribution \cite{Tsallis:52} describes the $p_{\rm T}$-spectra very well. In view of this, we have used a thermodynamically consistent 
form of Tsallis distribution function \cite{Cleymans:2011in}. The non-extensivity parameter $q$ gives the degree of deviation from equilibrium, where $q$ = 1 suggests the equilibrium condition (BG scenario). For higher charged particle multiplicity, the $q$ value tends to 1, which is an indication that in that regime, the system has most probably attained thermal equilibrium. By fitting Tsallis distribution function to the $p_{\rm T}$-spectra of the particles, the parameters $q$ and temperature $T$ are extracted \cite{Tsallis:2003vv}, which are then used to find the particle multiplicities in the condensate. The present formalism is motivated by the experimental $p_{\rm T}$-spectra and we use these information to further explore the possibility of a BEC
of pions in LHC $pp$ collisions.

In this paper, we have studied the possibility of pion condensation in high energy $pp$ collision systems at LHC energy of $\sqrt {s}$ = 7 TeV. We have also studied how the critical temperature changes with the change in the non-extensive parameter $q$. The section \ref{formulation} briefly gives the formulation for estimating the particle multiplicities in the condensate and the critical temperature for BEC formation. In section \ref{res}, we discuss about our findings and finally in section \ref{sum}, we have summarized our findings.

\section{Formulation}
\label{formulation}
Before introducing the non-extensivity into the formalism, let's start with a general distribution function of Bose-Einstein 
statistics, which is given as \cite{KHuang},
\begin{equation}
\label{eq1}
f = \frac {1}{exp(\frac{E - \mu}{T}) - 1}.
\end{equation}

By using the above formula, we can calculate the particle multiplicities by the equation \cite{Begun:2015ifa},
\begin{equation}
\label{eq1}
%N =   \int_{0}^{\infty}\frac{d^3xd^3p}{h^3}  \frac {g}{exp\bigg(\frac{\sqrt{p^2 + m^2} - \mu}{T}\bigg) - 1}\nonumber\\
N =   \int\frac{d^3xd^3p}{h^3}  \frac {g}{exp\bigg(\frac{\sqrt{p^2 + m^2} - \mu}{T}\bigg) - 1}\nonumber\\
\end{equation}
\begin{equation}
\label{eq2}
% \simeq  V \int_{0}^{\infty} \frac{d^3p}{(2\pi)^3} \frac {g}{exp\bigg(\frac{\sqrt{p^2 + m^2} - \mu}{T}\bigg) - 1},
 \simeq  V \int \frac{d^3p}{(2\pi)^3} \frac {g}{exp\bigg(\frac{\sqrt{p^2 + m^2} - \mu}{T}\bigg) - 1},
\end{equation}
where $g$ is the degeneracy of the particle, $p$ is the momentum, $m$ is the mass of the particle, $T$ is the temperature of the system and $\mu$ is the chemical potential. The integral over the space co-ordinates gives us the volume of the system $V$.

For p $\to$ 0, if $\mu \to m$, the integrand in first line of Eq. \ref{eq2} stays constant. This is because the singularity of the denominator is canceled by the integration measure $d^{3}\it {p}$. But in the sum over quantum levels, the first term becomes infinite for $\mu \to m$ at p $= 0$, i.e.,
\begin{equation}
 N_{\rm condensation} \simeq  \frac {g}{ exp(\frac{m - \mu}{T}) - 1} \to \infty ~\rm {for} ~\mu \to m
  \end{equation}
  
 Thus, when $\mu \to m$, which marks the onset of BEC, integration of Eq. \ref{eq2} should begin for p $>$ 0 and summation over low momentum should be kept. In the thermodynamic limit \cite{Begun:2008hq}, $V \to \infty$, one can write Eq. \ref{eq2} with separate terms for $p$ = 0 and $p$ $>$ 0, as:
\begin{equation}
 %N \simeq  \frac {g}{ exp(\frac{m - \mu}{T}) - 1} + V \int_{0}^{\infty} \frac{d^3p}{(2\pi)^3} \frac {g}{exp\bigg(\frac{\sqrt{p^2 + m^2} - \mu}{T}\bigg) - 1}\nonumber\\
 N \simeq  \frac {g}{ exp(\frac{m - \mu}{T}) - 1} + V \int \frac{d^3p}{(2\pi)^3} \frac {g}{exp\bigg(\frac{\sqrt{p^2 + m^2} - \mu}{T}\bigg) - 1}\nonumber\\
 \end{equation}
 \begin{equation}
 \label{eq3}
\Rightarrow N_{\rm total} = N_{\rm condensation} + N_{\rm excited}.
\end{equation}
When the chemical potential approaches to the value of the mass of the particle, $\mu \to m$, the $N_{\rm condensation}$ becomes dominant and tends to $N_{\rm total}$ at $\mu = m$. However at LHC $pp$ collisions, the chemical potential is very small and can be taken as zero. 

By taking Tsallis non-extensivity~\cite{Cleymans:2014woa,wilk1} into account, the BE-distribution function changes to,
\begin{equation}
\label{eq4}
f = \frac {1}{exp_{q}(\frac{E - \mu}{T}) - 1}.
\end{equation}
%Here, the non-extensive exponential function is defined as,
%\begin{equation}
%\label{eq5}
%exp_{q}(x) = [1 + (q-1)x]^{1/q-1}
%\end{equation}
%%============new lines=======
where, 
\begin{equation}
\label{expq}
\exp_{q}(x) \equiv
  \begin{cases}
    [1+(q-1)x]^{\frac{1}{q-1}}       & \quad \text{if }  x > 0\\
    [1+(1-q)x]^{\frac{1}{1-q}}      & \quad \text{if }  x \le 0\\
  \end{cases}
\end{equation}
where $x = (E-\mu)/T$, $E$ is the energy of the particle given by $E=\sqrt{p^2+m^2}$. 
%$p$ and $m$ being the  momentum and mass of the particle, respectively. 
It is worth noting that in the  limit, $q \rightarrow$ 1, Eq. \eqref{expq} reduces to the standard 
exponential function i.e., Maxwell-Boltzmann distribution function,
\begin{eqnarray*}
\lim_{q \to 1} \exp_q(x) \rightarrow \exp(x).
\end{eqnarray*}
The Tsallis parameter, $T$ and $q$ appearing in Eq.~\ref{expq} are extracted from the $p_{\rm T}$-spectra of the particle by using Tsallis distribution as a fitting function.

%%======================

Therefore, Eq.\ref{eq3} in the context of non-extensivity in its thermodynamically consistent form as shown in Ref~\cite{Cleymans:2011in} becomes,
\begin{equation}
 N \simeq  \frac {g}{ \bigg[exp_{q}(\frac{m - \mu}{T}) - 1\bigg]^q} + \nonumber\\
  \end{equation}
  \begin{equation}
\label{eq6}
% V \int_{0}^{\infty} \frac{d^3p}{(2\pi)^3} \frac {g}{\bigg[exp_{q}\bigg(\frac{\sqrt{p^2 + m^2} - \mu}{T}\bigg) - 1\bigg]^q}.
  V \int \frac{d^3p}{(2\pi)^3} \frac {g}{\bigg[exp_{q}\bigg(\frac{\sqrt{p^2 + m^2} - \mu}{T}\bigg) - 1\bigg]^q}.
 \end{equation} 

As per Ref~\cite{Begun:2006gj}, the formula for the critical temperature in the ultra-relativistic limits is given by,
\begin{equation}
\label{eq7}
T_{\rm c} = 1.4 \times \rho^{1/3},
 \end{equation}
where, $\rho$ is the pion number density of the system, given by the formula  within Tsallis statistics as follows~\cite{Cleymans:2012ya},
\begin{equation}
\label{eq8}
%\rho = g \int \frac{d^{3}p}{(2\pi)^3}\bigg[1 + (q-1)\frac{E- \mu}{T} \bigg]^\frac{-q}{q-1} ,
\rho = g \int \frac{d^{3}p}{(2\pi)^3}\bigg[[1 + (q-1)\frac{E- \mu}{T} ]^\frac{1}{q-1} -1\bigg]^{-q}.
 \end{equation}

One can see from Eqns. \ref{eq7} and \ref{eq8} the critical temperature is also dependent on number density, $\rho$. The higher is the number density, higher is the critical temperature of the system as shown in Ref.~\cite{Begun:2006gj}. Further in high energy collisions, the values of 
the $q$-parameter can vary between 1 to 11/9~\cite{beck}, which also contributes to the estimation of $T_{\rm c}$ taking the system dynamics
into account.

\section{Results and Discussion}
\label{res}

We have used Eq.\ref{eq6} to estimate the particle multiplicities in the excited states and the condensate. We have taken certain values of temperature and have estimated the multiplicities at particular $q$ values. For the sake of simplicity, here we have taken a constant value for the volume of the system with the system radius of 1.2 fm, which can be approximately taken as the chemical freeze-out radius in $pp$ collision systems. This is a reasonable assumption as the HBT radii measurement from $pp$ collisions at the LHC gives the radius range of 0.8 - 1.6 fm~\cite{Aamodt:2011kd}. 

\begin{figure}[ht!]
\begin{center}
\includegraphics[scale = 0.434]{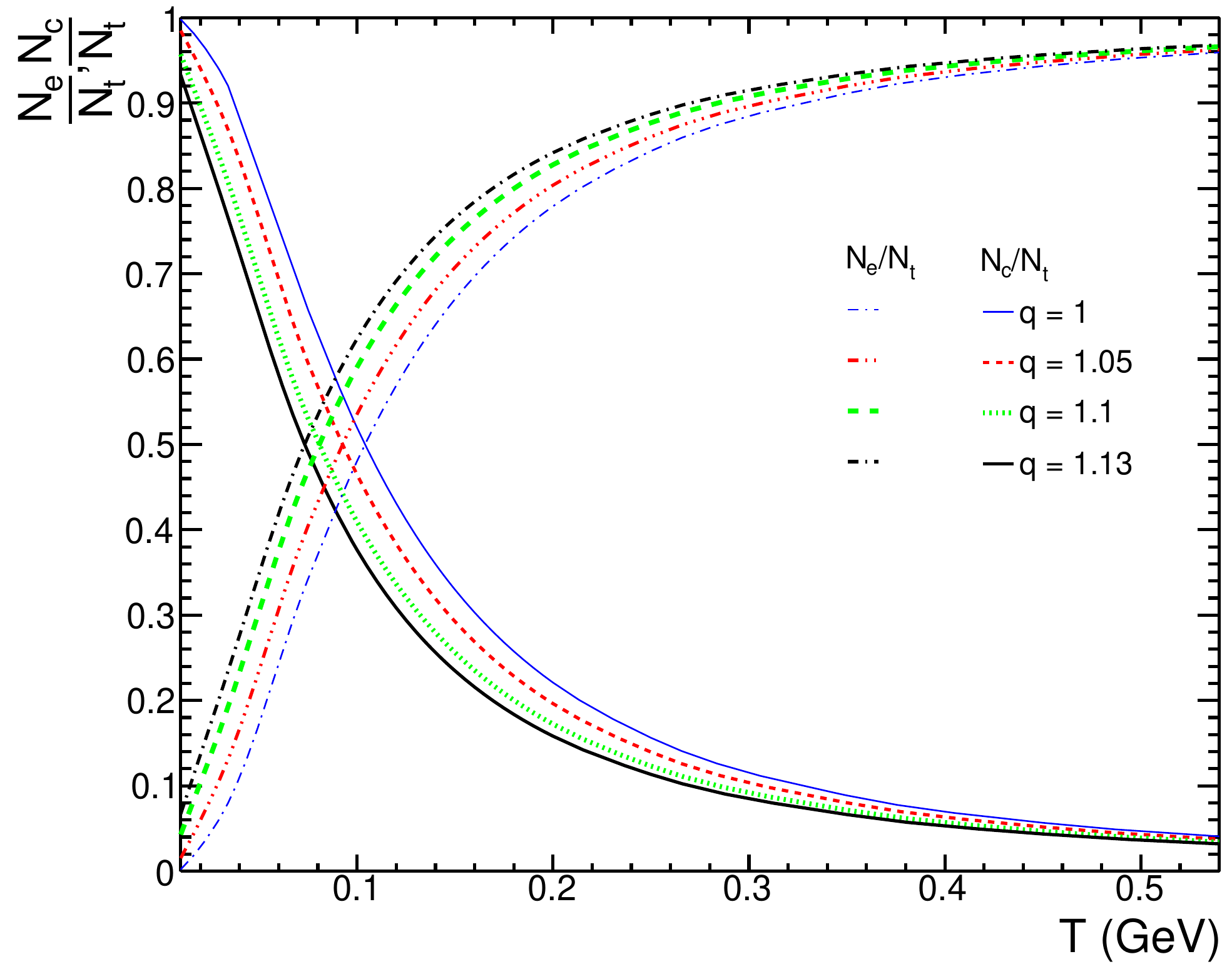}
\caption{(Color online) Ratios of number of particles in the condensates and number of particles in the excited states with respect to the total number of particles as a function of temperature for a pion gas for different $q$-values}
\label{fig1}
\end{center}
\end{figure}

In fig. \ref{fig1}  we have plotted the ratios of $N_{\rm condensation}$ to $N_{\rm total}$ $(\rm {N}_{c}/\rm {N}_{t})$ and $N_{\rm excited}$ to $N_{\rm total}$ $(\rm {N}_{e}/\rm {N}_{t})$ as a function of temperature using Eq. \ref{eq6}. We observe that at high temperatures, the particle multiplicities in the excited states is dominant as compared to that in the condensate. As the temperature decreases, the particle multiplicities in the condensate start to increase. After a certain critical temperature, the number of particles in the condensate becomes dominant. We also clearly see that the critical temperature is highly $q$-dependent. For BE statistics without using non-extensivity ($q$=1), we find that the critical temperature is the highest at around 105 MeV. However, as we increase the $q$-value, the critical temperature starts to decrease. For $q$ = 1.13, we have the lowest critical temperature at around 75 MeV. This means that for systems which are near equilibrium, the critical temperature will be higher, whereas for the systems which are away from equilibrium, the critical temperatures will be relatively lower. This is a really interesting finding since we know that for the highest multiplicity of $\pi^{\pm}$  in the $pp$ collisions at $\sqrt {s}$ = 7 TeV, the kinetic freeze-out temperature obtained after fitting a thermodynamically consistent Tsallis distribution function to its $p_{\rm T}$-spectra is around 93 MeV~\cite{Khuntia:2018znt}. This may indicate that there is a possibility that we may see BEC in $pp$ collisions at LHC energies.

\begin{figure}[ht!]
\begin{center}
\includegraphics[scale = 0.436]{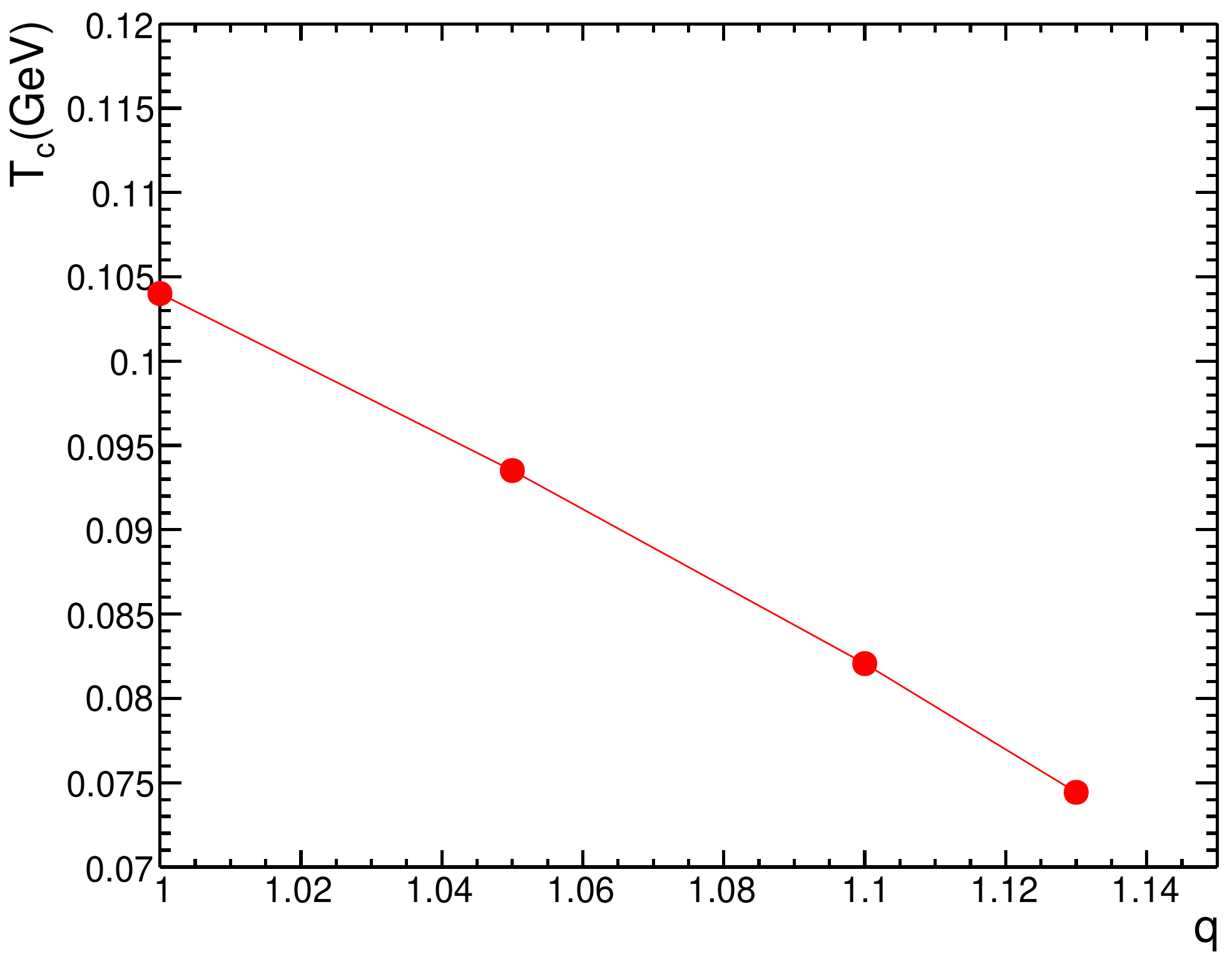}
\caption{(Color online) Critical temperature as a function of non-extensive parameter $q$}
\label{fig2}
\end{center}
\end{figure}

In fig.\ref{fig2} we have plotted the critical temperature ($T_c$) as a function of the non-extensive parameter $q$ extracting the values from the crossing points seen in fig.\ref{fig1}. We see that for higher values of $q$, which means when the system is far away from equilibrium, the critical temperature for BEC is lower. When, the system is at equilibrium ($q$ = 1), the critical temperature is the highest at around 105 MeV. An important thing to note here is that for a system which has acquired equilibrium, the critical temperature will be higher, and it may not be possible for the pion gas system to reach that temperature in the freeze-out. So there is a distinct possibility that if a system is in thermal equilibrium, it may not show BEC~\cite{Greiner:1993jn}.

\begin{figure}[ht!]
\begin{center}
\includegraphics[scale = 0.438]{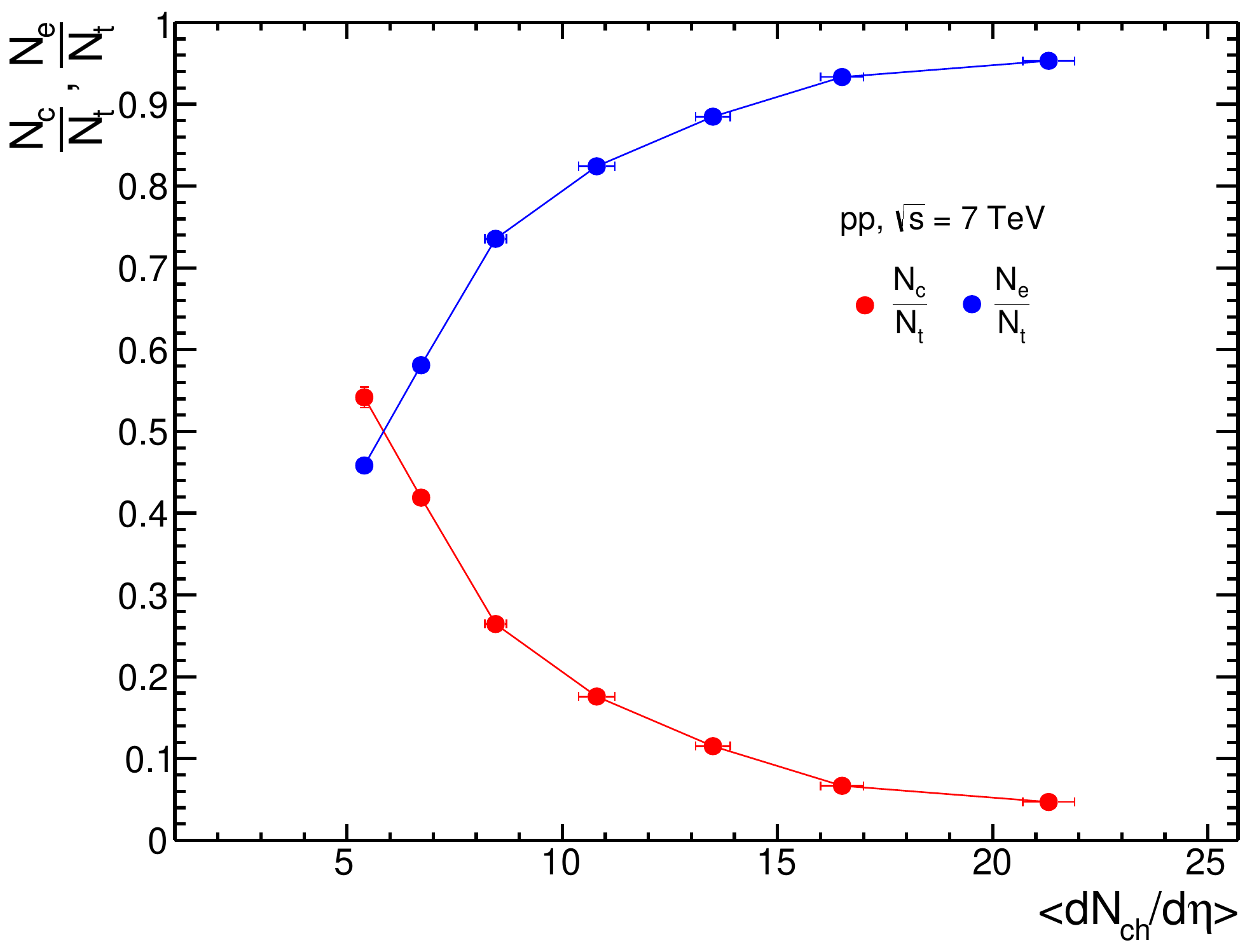}
\caption{(Color online) Ratios of number of particles in the condensates and number of particles in the excited states with respect to the total number of particles
as a function of charged particle multiplicity for $pp$ collisions at $\sqrt{s}$ = 7 TeV in ALICE at the Large Hadron Collider}
\label{fig4}
\end{center}
\end{figure}

In fig. \ref{fig4} we have plotted the ratios of $(\rm {N}_{c}/\rm {N}_{t})$ and $(\rm {N}_{e}/\rm {N}_{t})$ for $pp$ collisions at $\sqrt{s}$ = 7 TeV using ALICE data \cite{Acharya:2018orn} as a function of charged particle multiplicity at midrapidity, which uses Eq. \ref{eq6}. T and $q$ values used in Eq. \ref{eq6} are taken from Ref. \cite{Khuntia:2018znt}, where the $p_{\rm T}$-spectra of produced identified particles in a differential freeze-out scenario are fitted by Tsallis distribution function. To have a quantitative estimation
of the volume of the system at the kinetic freeze-out, we consider the chemical freeze-out radius from Ref~\cite{Sharma:2018jqf} and as the system is an expanding one, we add hadronic phase lifetime from our earlier studies \cite{Sahu:2019tch} multiplied with speed of light in vacuum.  The radii obtained by this method is quite comparable with 
that one obtains from the femtoscopy analysis~\cite{Aamodt:2011kd}. It is worth mentioning here that this assumption ensures maximum radii of the system. We observe that at high charged particle multiplicity which corresponds to higher kinetic freeze-out temperature, about 95\% of the particles are in the excited states and about 5\% of the particles occupy the ground state. With the decrease in $\langle dN_{\rm ch}/d\eta \rangle$, we observe that the $N_{\rm excited}$ to $N_{\rm total}$ ratio  decreases while the $N_{\rm condensation}$ to $N_{\rm total}$ ratio increases. At about $\langle dN_{\rm ch}/d\eta \rangle$ $\simeq$ 6, which corresponds to 78 MeV temperature~\cite{Khuntia:2018znt}, we observe a transition. This $\langle dN_{\rm ch}/d\eta \rangle$ can be considered as the critical charged particle multiplicity for $pp$ collisions at $\sqrt {s}$ = 7 TeV at LHC, below which the number of particles in the condensate is higher than the excited states. At the lowest charged particle multiplicity, we observe that the particle multiplicity in the condensate is dominant over the particles in the excited states. This is an interesting finding given that at low charged particle multiplicity the number density, the volume and the temperature of the system are relatively lower as compared to the systems at high charged particle multiplicities. Similar results were observed by Begun et.al. where they found number of particles in the condensate relative to the number of particles in the excited state to be higher in peripheral collisions (low multiplicity events) than central collisions (high multiplicity events) \cite{Begun:2015ifa}.

\begin{figure}[ht!]
\begin{center}
\includegraphics[scale = 0.43]{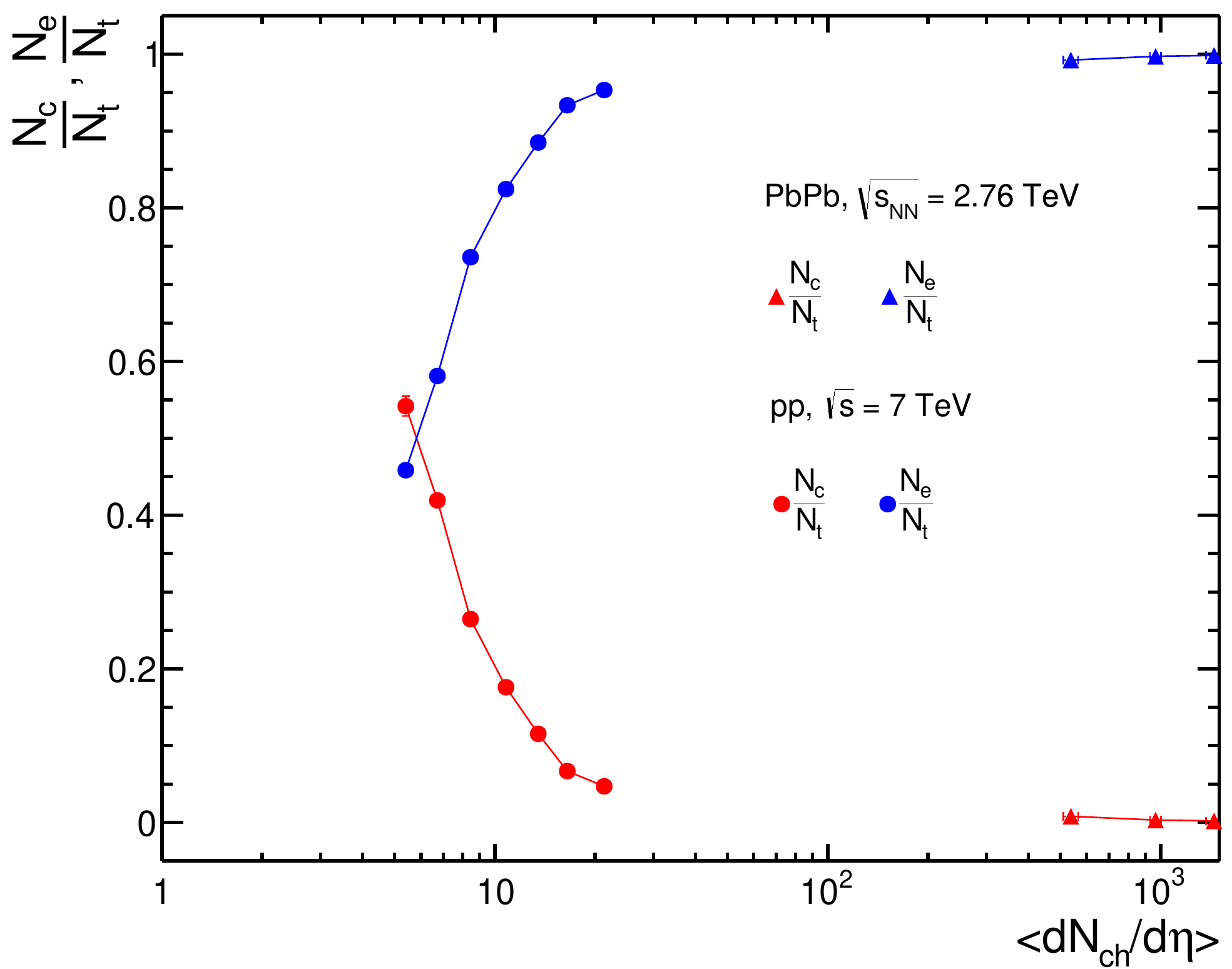}
\caption{(Color online) Ratios of number of particles in the condensates and number of particles in the excited states with respect to the total number of particles as a function of charged particle multiplicity for $pp$ collisions at $\sqrt{s}$ = 7 TeV and Pb-Pb collisions at $\sqrt{s_{\rm NN}}$ = 2.76 in ALICE at the Large Hadron Collider}
\label{fig5}
\end{center}
\end{figure}

To have a comparison with the systems produced in heavy-ion collisions, we have considered the possibility of pion condensation in the Pb-Pb collisions at $\sqrt{s_{\rm NN}}$ = 2.76 TeV and contrasted with the results from Fig~\ref{fig4}. For the former system, we have taken the $T$ and $q$ parameters from~\cite{Azmi:2019irb} which have been extracted by fitting a thermodynamically consistent Tsallis distribution function to the $p_{\rm T}$-spectra. From fig.\ref{fig5}, we observe that for Pb-Pb collision systems, the number of particles in the excited states are higher at higher charged particle multiplicity. In this region, the temperature of the system is very high and the non-extensive parameter $q$ has lower value, and almost tends to 1. This suggests that the system is in a state of thermodynamic equilibrium. As the charged particle multiplicity decreases, which in turn means the temperature also decreases and the $q$ value increases, we observe that particles start accumulating in the ground state. The number of particles in the condensate begins to increase from almost zero in the highest charged particle multiplicity to higher value in the most peripheral Pb-Pb collisions as shown in Ref.~\cite{Begun:2015ifa}. When we go to $pp$ collisions, we observe a smooth transition from collision species following a monotonic behaviour with final state multiplicity of the system. The figure shows that at around $\langle dN_{\rm ch}/d\eta \rangle$ $\simeq$ 6, we may observe BEC, regardless of the collision systems. 

\section{Summary}
\label{sum}
We have made an attempt to study the possible formation of Bose-Einstein Condensates in the hadronic collisions at the TeV
energies at the Large Hadron Collider, using the information from identified particle spectra. Motivated by the fact that a 
thermodynamically consistent form of Tsallis non-extensive distribution function gives a better description of the particle
transverse momentum spectra, we have extended the standard formalism of BEC to the domain of non-extensivity. Taking 
final state multiplicity dependence of the identified particle spectra and hence the freeze-out parameters like temperature and non-extensive parameter, $q$ for hadronic and heavy-ion collisions, we have studied various aspects of BEC. 
 
In summary, 

\begin{itemize}
\item[$\bullet$] In the low-multiplicity $pp$ collisions, where the systems are away from equilibrium ($q \neq 1$), the probability
of formation of BEC is seen to be higher. 

\item[$\bullet$] The critical temperature for the BEC to occur, depends on the degree of non-equilibrium in the system. This is
observed from the dependency of the number of pions in the condensates on the non-extensive parameter $q$. For systems away from
equilibrium, the values of critical temperature are lower.

\item[$\bullet$] We observe a threshold in the final state charged-particle multiplicity, {\it i.e.} $\langle dN_{\rm ch}/d\eta \rangle$ $\simeq$ 6, which corresponds to an event class with freeze-out temperature of around 78 MeV.  In this domain of temperature and particle density, we envisage a possibility of BEC occurring in $pp$ collision at $\sqrt {s}$ = 7 TeV energy at the Large Hadron Collider.

%\end{enumerate}
\end{itemize}
 
 As a part of the outlook of the present work, it should be mentioned here explicitly that a Bose-Einstein condensate of strongly 
 interacting matter, having an extremely large temperature as compared to the conventional systems, consisting of charged 
 particles would be highly interesting in the coming future, if seen experimentally. The low transverse momentum reach of the
 TeV scale detector systems after the upgrade programs at the LHC will be much better at enabling such a measurement in future.
  
\section{Acknowledgement} 
 This research work has been carried out under the financial supports from DAE-BRNS, Government of India, Project No. 58/14/29/2019-BRNS of Raghunath Sahoo.

 \end{document}